\documentstyle[preprint,eqsecnum,aps,amsfonts,amssymb,psfig,epsfig,prb]{revtex} 

\begin{document}


\def\reff#1{(\ref{#1})}
\newcommand{\be}{\begin{equation}}
\newcommand{\ee}{\end{equation}}
\newcommand{\<}{\langle}
\renewcommand{\>}{\rangle}

\def\spose#1{\hbox to 0pt{#1\hss}}
\def\ltapprox{\mathrel{\spose{\lower 3pt\hbox{$\mathchar"218$}}
 \raise 2.0pt\hbox{$\mathchar"13C$}}}
\def\gtapprox{\mathrel{\spose{\lower 3pt\hbox{$\mathchar"218$}}
 \raise 2.0pt\hbox{$\mathchar"13E$}}}

\def\bsigma{\mbox{\protect\boldmath $\sigma$}}
\def\bpi{\mbox{\protect\boldmath $\pi$}}
\def\smfrac#1#2{{\textstyle\frac{#1}{#2}}}
\def\smhalf{ {\smfrac{1}{2}} }

\newcommand{\re}{\mathop{\rm Re}\nolimits}
\newcommand{\im}{\mathop{\rm Im}\nolimits}
\newcommand{\tr}{\mathop{\rm tr}\nolimits}
\newcommand{\fr}{\frac}
\newcommand{\diti}{\frac{\mathrm{d}^2t}{(2 \pi)^2}}

\def\Z{{\mathbb Z}}
\def\R{{\mathbb R}}
\def\C{{\mathbb C}}

\title{Polymer size in dilute solutions in the good-solvent regime}

\author{ Sergio Caracciolo, Bortolo Matteo Mognetti}
\address{Dipartimento di Fisica and INFN -- Sezione di Milano I
  Universit\`a degli Studi di Milano \\
  Via Celoria 16, I-20133 Milano, Italy \\  
  e-mail: {\tt Sergio.Caracciolo@sns.it},
          {\tt Bortolo.Mognetti@mi.infn.it}}
\author{ Andrea Pelissetto }
\address{
  Dipartimento di Fisica and INFN -- Sezione di Roma I  \\
  Universit\`a degli Studi di Roma ``La Sapienza'' \\
  P.le A. Moro 2, I-00185 Roma, Italy \\
  e-mail: {\tt Andrea.Pelissetto@roma1.infn.it }}

\maketitle
\thispagestyle{empty}   

\begin{abstract}

We determine the density expansion of the radius of gyration, of the 
hydrodynamic radius, and of the end-to-end distance for a monodisperse polymer 
solution in good-solvent conditions. We consider the scaling limit 
(large degree of polymerization), including the leading scaling 
corrections. Using the expected large-concentration behavior,
we extrapolate these low-density expansions outside the
dilute regime, obtaining a prediction for the radii for any concentration in
the semidilute region. For the radius of gyration, 
comparison with field-theoretical predictions 
shows that the relative error should be at most 5\% 
in the limit of very large polymer concentrations.

\bigskip 



\end{abstract}

\clearpage

\section{Introduction}

For sufficiently high molecular weights, dilute and semidilute 
polymer solutions under good-solvent conditions exhibit a universal 
scaling behavior, as predicted by the renormalization 
group.\cite{deGennes-72,deGennes-79,Freed-87,dCJ-book,Schaefer-99}
For instance, if $R$ is some typical dimension of the polymer, 
then $R\approx a N^\nu$, where $N$ is the degree of 
polymerization, $a$ is a constant that depends on 
density, temperature, and chemical details, and 
$\nu$ a universal exponent, $\nu \approx 0.5876$
(Ref.~\CITE{bestnu}). From an experimental point of view, two 
quantities $R$ are easily accessible in the dilute regime.
The radius of gyration $R_g$ is determined in small-angle 
scattering experiments. For small wave numbers $q$, the static 
form factor behaves as
\be
F(q) = {1\over N^2} \sum_{\alpha\beta}
  \langle e^{i{\mathbf q}\cdot ({\mathbf r}_\alpha - {\mathbf r}_\beta)}
   \rangle \approx 
   1 - {q^2\over 3} R^2_g + O(q^4),
\ee
that allows the determination of $R_g$. The second quantity of interest is 
the hydrodynamic radius $R_H$, which is related to the chain diffusion constant
$D$ that can be measured in inelastic scattering experiments:\cite{DE-book}
\be
D = {k_B T\over 6 \pi\eta R_H} ,
\ee
where $\eta$ is the viscosity. In theoretical work, 
another quantity is often used, the end-to-end distance $R_e$, which can, 
in principle, be accessible in experiments with polymers with 
tagged endpoints.\cite{dCJ-book}

In this paper we consider the density dependence of these three quantities.
In general, if $R$ is one of the quantities defined above or any
functions of them, 
in the dilute regime we can write 
\be
R(c) = \hat{R} [1 + R_1 c + R_2 c^2 + O(c^3) ],
\ee
where $c$ is the polymer number density and $\hat{R} \equiv  R(c = 0)$. 
The coefficients $R_1$, $R_2$, etc. depend on chemical details. 
However, as $N\to \infty$ the ratios $S_n\equiv R_n/\hat{R}_g^{3n}$ 
approach universal constants $S_n^*$ 
that are independent of temperature---as long 
as the polymer is in the good-solvent regime---and chemical properties.
Thus, if $N$ is large (scaling regime) we can write the universal 
expression
\be
{R(c)\over \hat{R}} = 1 + S_1^* c\hat{R}_g^{3} + S_2^* (c\hat{R}_g^{3})^2 + 
   O[(c\hat{R}_g^{3})^3].
\ee
In many cases, polymers are not long enough to be in the scaling regime
and thus it is important to include scaling corrections.  As predicted 
by the renormalization group and extensively verified numerically, 
for $N$ large but finite we have for the expansion coefficients 
of quantities related to the end-to-end distance and to the 
radius of gyration
\begin{equation}
S_{n} = {R_n \over \hat{R}^{3n}_g} \approx 
     S_n^*\left(1 + k_2 s_{n} N^{-\Delta}\right),
\label{corrections}
\end{equation}
where $\Delta$ is a universal exponent whose best estimate 
is\cite{BN-97} $\Delta = 0.515\pm 0.007^{+0.010}_{-0.000}$.
The constants $S_{n}^*$ and $s_n$ are universal. 
The constant $k_2$ encodes all chemical details and nonuniversal
properties of the solution---temperature, for instance. It can be determined
by considering the scaling corrections to some renormalization-group
invariant polymer quantity. We use the interpenetration radius 
$\Psi$ that can be determined from the osmotic pressure and 
represents one of the most easily experimentally accessible 
quantities.\cite{footPsi} In the large-$N$ limit it behaves as
\begin{equation}
\Psi \approx \Psi^* \left(1 + k_2 N^{-\Delta}\right).
\label{Psi-exp}
\end{equation}
This relation defines the constant $k_2$. 
The constant $\Psi^*$ is universal and known with very good 
precision:\cite{CMP-viriale}
$\Psi^* = 0.24693\pm 0.00013$. 
Other estimates are reported in Ref.~\CITE{PV-review}.
Universal ratios related to the hydrodynamic radius scale differently because 
of the presence of corrections proportional to $N^{1-\nu}$ 
(Ref.~\CITE{DRSK-02}). Note finally that, given two different quantities
$R_1$ and $R_2$ that characterize the polymer size, their zero-density ratio 
$\hat{R}_1/\hat{R}_2$ approaches a universal constant 
$\hat{R}_{12}^*$ for $N\to\infty$. 

For the purpose  of determining the universal constants defined above 
we perform a Monte Carlo simulation
of the lattice Domb-Joyce model,\cite{DJ-72} considering walks 
of length varying between 100 and 8000 and three different 
penalties for the self-intersections. We are able to obtain 
the leading density correction with a relative precision of 0.5\% for 
$R_g$ and $R_e$ and of 5\% for $R_H$. We also estimate the second 
density correction, although with limited precision. 
These results show that density corrections
are very small below the overlap concentration 
$c^* \equiv  3/(4 \pi \hat{R}_g^3)$, and that a simple 
approximation that assumes a linear dependence on the density works 
quite well up to $c \approx c^*$. Once the virial expansion is
known, we can try to resum it to obtain an
interpolation formula that is valid in the whole semidilute regime.
In Ref.~\CITE{CMP-viriale} we have shown that 
a simple extrapolation of the virial expansion for the 
osmotic pressure that takes into account the 
expected large-concentration behavior works reasonably well.
Comparison with experimental data and with other theoretical 
predictions shows that the error is less than 1\% in the dilute region
and rises at most to 5-10\% deep in the semidilute regime. 
We use here the same strategy for the radii, determining simple expressions
that reproduce the virial expansion for $c\to0$ and have the correct 
large-density behavior. Comparison with other theoretical results 
shows that, at least for $R_g$, the interpolation formula has an error 
of at most 5\%. 

The paper is organized as follows. In Sec.~\ref{sec2} we derive the 
virial expansion for a generic quantity that depends on the structure of 
a single polymer, like the radii defined above. 
In Sec.~\ref{sec3} we define the 
model and the quantites we compute. 
In Sec.~\ref{sec4} we analyze the Monte Carlo results.
In Sec.~\ref{sec5} we present our conclusions and compare our results 
with those obtained in other approaches.

\section{Virial expansion for single-polymer properties} \label{sec2}

We wish now to derive the density expansion of $R(c)$, where $R$ is
a single-polymer property.
We consider $L$ polymers in a volume $V$ with 
configurational partition function
\begin{eqnarray}
Q_L&=& \int d\mu\, \exp\left(-\beta\sum_{i<j} V_{ij}^{\rm inter}\right)
\\
d\mu &\equiv& \left[\prod_{i\alpha} d^3{\mathbf x}_{\alpha i}\right] 
   \exp\left(-\beta\sum_{i} V_{i}^{\rm intra}\right),
\end{eqnarray}
where $V_{ij}^{\rm inter}$ is the sum of all terms of the Hamiltonian
that correspond to interactions of monomers belonging to two
different polymers $i$ and $j$, 
$V_{i}^{\rm intra}$ is the contribution due to interactions of monomers
belonging to the same polymer $i$, and ${\mathbf x}_{\alpha i}$ is 
the position of monomer $\alpha$ belonging to polymer $i$. 
As usual, we define the Mayer function
\begin{equation}
f_{ij} \equiv  \exp(- \beta V_{ij}^{\rm inter}) - 1.
\end{equation}
Since $R$ is a single-polymer property, we can assume 
that it depends only on the coordinates ${\mathbf x}_{\alpha 1}$
of the monomers of the first polymer.  Thus, we wish to compute the 
expansion of
\begin{equation}
\langle R \rangle = {1\over Q_L} \int d\mu\,  R(\{{\mathbf x}_{\alpha 1}\}) 
        \exp\left(-\beta\sum_{i<j} V_{ij}^{\rm inter}\right),
\end{equation}
for $c \to 0$.
The term linear in the number density $c \equiv L/V$ is easily derived:
\begin{eqnarray}
{Q_L\over Q_{0,L}} \langle R \rangle &=& \langle R \rangle_0 + 
   {L-1 \choose 2} \langle R f_{23} \rangle_0 + 
   (L-1) \langle R f_{12} \rangle_0 + \ldots
\\
{Q_L\over Q_{0,L}} &=& 1 + {L\choose 2} \langle f_{12} \rangle_0 + \ldots
\end{eqnarray}
where 
\begin{eqnarray}
Q_{0,L} \equiv  \int d\mu \qquad\qquad 
\langle {\cal O} \rangle_0 \equiv  {1\over Q_{0,L}} \int d\mu\,   {\cal O} .
\end{eqnarray}
It follows that, for $L,V\to\infty$ at fixed concentration, we have
\be
\langle R \rangle =  \hat{R} + c \left[J_1(R) - \hat{R} I_2\right],
\ee
where $\hat{R}$ is the zero-density value of $R$, and 
\begin{eqnarray}
I_2 &\equiv & \int d^3 {\mathbf r}_{12} \, 
       \langle f_{12} \rangle_{{\mathbf 0},{\mathbf r}_{12}} ,
\\
J_1(R)  &\equiv & \int d^3 {\mathbf r}_{12} \, 
       \langle R f_{12} \rangle_{{\mathbf 0},{\mathbf r}_{12}} \; .
\end{eqnarray}
Here $\langle \cdot \rangle_{{\mathbf 0},{\mathbf r}}$ indicates an average 
over two independent polymers such that the first one starts at the 
origin and the second starts in ${\mathbf r}$. In the mean values $R$ 
is a function of the coordinates of the first walk.

The correction of order $c^2$ can be worked out analogously. We define  
\begin{eqnarray}
I_3 &\equiv & 
    \int d^3{\mathbf r}_{12} d^3{\mathbf r}_{13}
     \left\langle f_{12}f_{13}f_{23}
     \right\rangle_{{\mathbf 0},{\mathbf r}_{12},{\mathbf r}_{13}} 
\\
T_1 &\equiv & \int d^3{\mathbf r}_{12} d^3{\mathbf r}_{13}
      \left\langle f_{12} f_{13} 
       \right\rangle_{{\mathbf 0},{\mathbf r}_{12},{\mathbf r}_{13}}
   -\left[ \int d^3{\mathbf r}_{12}
    \left\langle f_{12} \right\rangle_{{\mathbf 0},{\mathbf r}_{12}}\right]^2
\\
J_2(R) &\equiv & \int d^3{\mathbf r}_{12} d^3{\mathbf r}_{13}
      \left\langle R f_{12} f_{13} 
       \right\rangle_{{\mathbf 0},{\mathbf r}_{12},{\mathbf r}_{13}}
\\
J_3(R) &\equiv & \int d^3{\mathbf r}_{12} d^3{\mathbf r}_{13}
      \left\langle R f_{12} f_{23} 
       \right\rangle_{{\mathbf 0},{\mathbf r}_{12},{\mathbf r}_{13}}
\\
J_4(R) &\equiv & 
    \int d^3{\mathbf r}_{12} d^3{\mathbf r}_{13}
     \left\langle R f_{12}f_{13}f_{23}
     \right\rangle_{{\mathbf 0},{\mathbf r}_{12},{\mathbf r}_{13}} ,
\end{eqnarray}
where 
$\langle \cdot \rangle_{{\mathbf 0},{\mathbf r}_2,{\mathbf r}_3}$
refers to averages over three polymers, 
the first one starting in the origin, the second in 
${\mathbf r}_2$, etc. We obtain finally
\begin{eqnarray}
\langle R \rangle &=&  \hat{R} + c \left[J_1(R) - \hat{R} I_2\right] + 
\nonumber \\
&& {1\over2} c^2 \left[ J_2(R) + 2 J_3(R) + J_4(R) - 3 \hat{R} T_1 -
          \hat{R}  I_3 - 4 I_2 J_1(R) + \hat{R} I_2^2\right] + O(c^3).
\label{expanR1}
\end{eqnarray}

In the following we shall consider a lattice model for polymers. 
In this case, the previous expressions must be trivially modified, 
replacing each integral with the corresponding sum over all 
lattice points. 

\section{Model and observables} \label{sec3}

Since we are interested in computing universal quantities,
we can use any model that captures the basic 
polymer properties. For computational convenience we shall consider 
a lattice model.  A polymer of length
$N$ is modelled by a random walk 
$\{{\mathbf r}_0,{\mathbf r}_1,\ldots,{\mathbf r}_N\}$ with 
$|{\mathbf r}_\alpha-{\mathbf r}_{\alpha+1}|=1$ on a cubic lattice. 
To each walk we associate a Boltzmann factor
\begin{equation}
e^{-\beta H} = e^{-w\sigma},\qquad\qquad
\sigma = \sum_{0\le \alpha < \beta \le N} 
   \delta_{{\mathbf r}_\alpha,{\mathbf r}_\beta},
\end{equation}
with $w > 0$. The factor $\sigma$ counts how many 
self-intersections are present in the walk. This model is similar
to the standard self-avoiding walk (SAW) model in which polymers are 
modelled by random walks in which self-intersections are forbidden. 
The SAW model is obtained for $w = +\infty$. For finite positive $w$ 
self-intersections are possible although energetically penalized. 
For any positive $w$, this model---hereafter we will refer to it
as Domb-Joyce (DJ) model---has the same scaling limit of the 
SAW model\cite{DJ-72} and thus allows us to compute the 
universal scaling functions that are relevant for polymer solutions.
The DJ model has been extensively studied numerically in Ref.~\CITE{BN-97}.
There, it was also shown that there is a particular value of $w$,
$w^*$, for which corrections to scaling 
with exponent $\Delta$ vanish: the nonuniversal constant 
$k_2$ is zero for this particular value of $w$. Thus, simulations
at $w = w^*$ are particularly convenient since the scaling limit
can be observed for smaller values of $N$. In Ref.~\CITE{BN-97} 
$w^*$ was shown to be approximately equal to 0.505838 ($e^{-w^*} = 0.603$),
while in Ref.~\CITE{CMP-viriale} it was found $w^*\approx 0.47\pm 0.02$. 

The DJ model can be efficiently simulated by using the pivot 
algorithm.\cite{Lal,MacDonald,Madras-Sokal,Sokal-95b}
For the SAW an efficient implementation is discussed in 
Ref.~\CITE{Kennedy-02}. The extension to the DJ model is straightforward,
the changes in energy being taken into account by means of 
Metropolis test. Such a step should be included carefully
in order not to loose the good scaling behavior of the CPU time 
for attempted move. We use here the implementation discussed in 
Ref.~\CITE{CPP-94}.

We shall be interested in the quantities that characterize the size of the 
polymer and have been defined in the introduction. 
In the lattice model they are defined as follows:
\begin{eqnarray}
R^2_g &=& {1\over 2(N+1)^2} \left\langle \sum_{\alpha\beta} 
   ({\mathbf r}_{\alpha} - {\mathbf r}_\beta)^2\right\rangle, 
\\
{1\over R_H} &=& {1\over (N+1)^2} \left\langle 
     \sum_{\alpha\beta:{\mathbf r}_{\alpha}\not={\mathbf r}_\beta}
   {1\over |{\mathbf r}_{\alpha} - {\mathbf r}_\beta|} \right\rangle, 
\\
R^2_e &=& \langle ({\mathbf r}_{0} - {\mathbf r}_N)^2\rangle\; .
\end{eqnarray}
The expansion coefficients $S_n$ have been determined
by using the expressions reported in Sec.~\ref{sec2} with
\begin{eqnarray}
&& f_{ij} = e^{-\beta V_{ij}} - 1 = e^{-w\sigma_{ij}} - 1
\\
&& \sigma_{ij} = \sum_{\alpha\beta} 
   \delta_{{\mathbf r}_{\alpha i},{\mathbf r}_{\beta j}}.
\end{eqnarray}
Here ${\mathbf r}_{\alpha i}$ is the position of monomer $\alpha$ of polymer
$i$. To compute the corresponding lattice sums, 
we use a simple generalization of the hit-or-miss 
algorithm discussed in Ref.~\CITE{LMS-95} and in the appendix 
of Ref.~\CITE{CMP-viriale}. 

In the dilute limit we write
\begin{eqnarray}
{R^2_g\over \hat{R}_g^2} &=& 1 + S_{1,g} (c\hat{R}_g^3) + 
       S_{2,g} (c\hat{R}_g^3)^2 + \cdots 
\\
{R^2_e\over \hat{R}_e^2} &=& 1 + S_{1,e} (c\hat{R}_g^3) + 
       S_{2,e} (c\hat{R}_g^3)^2 + \cdots
\end{eqnarray}
The coefficients $S_{n,\#}$ have a finite limit for  $N\to \infty$.
Including the leading scaling correction, in the scaling limit
they behave as follows:
\begin{equation}
S_{n,\#}(N) = S_{n,\#}^*\left(1 + {k_2 s_{n,\#} \over N^\Delta} + 
   \cdots \right),
\end{equation}
where the constants $S_{n,\#}^*$ and $s_{n,\#}$ are universal. The factor 
$k_2$ depends instead on chemical details, temperature, etc. and is fixed 
by the large-$N$ behavior (\ref{Psi-exp}) of the interpenetration radius
$\Psi$.\cite{footPsi} 

Beside the coefficients parametrizing the density dependence of the radii,
we can also consider the large-$N$ behavior of the radii that we parametrize
as 
\begin{eqnarray}
\hat{R}^2_g &=& a_g N^{2\nu} \left(1 + k_2 r_{g} N^{-\Delta} + \cdots\right), \\
\hat{R}^2_e &=& a_e N^{2\nu} \left(1 + k_2 r_{e} N^{-\Delta} + \cdots\right). 
\end{eqnarray}
The constants $r_g$ and $r_e$ are universal as well as the ratio
$A^*_{ge} \equiv a_g/a_e$. This quantity can also be determined directly 
by considering 
\begin{equation}
A_{ge} = {\hat{R}^2_g\over \hat{R}^2_e} \approx 
     A_{ge}^* \left(1 + k_2 r_{ge} N^{-\Delta} + \cdots \right),
\end{equation}
where $r_{ge} = r_g - r_e$. 

Quantities involving the hydrodynamic radius must be analyzed 
differently, due to the appearance of new nonuniversal corrections proportional
to $N^{\nu-1}$. For instance,  
as discussed in Ref.~\CITE{DRSK-02}, $1/\hat{R}_H$, which is the 
quantity measured in MC calculations, scales as 
\be
{1\over \hat{R}_H} = {\alpha_1\over N^\nu} (1 + a_1 N^{-\Delta}) + {b\over N} + \cdots = 
         {\alpha_1\over N^\nu} \left(1 + {b\over \alpha_1} N^{\nu - 1} + 
            a_1 N^{-\Delta} + \cdots \right)\; .
\ee
The analytic term gives rise to a new correction proportional
to $N^{\nu-1} = N^{-0.412}$. It decays slower than the standard
one proportional to $N^{-\Delta}=N^{-0.5}$ and does not satisfy
any universality property. We also consider the density expansion 
\begin{eqnarray}
{\hat{R}_H\over R_H} &=& 1 + S_{1,H} (c\hat{R}_g^3) + 
       S_{2,H} (c\hat{R}_g^3)^2 + \cdots 
\end{eqnarray}
and the ratio 
\begin{equation}
A_{gH} \equiv  {\hat{R}_g\over \hat{R}_H}.
\end{equation}
For $N\to \infty$, $S_{n,H}$ and $A_{g,H}$ converge to 
universal constant $S_{n,H}^*$ and $A_{gH}^*$.
Note that it is more convenient to consider $\hat{R}_g/\hat{R}_H$ than 
$\hat{R}_H/\hat{R}_g$,
since the latter quantity has additional corrections
proportional to $N^{2 \nu - 2}= N^{-0.825}$, etc.

\section{Analysis of the Monte Carlo data} \label{sec4}

We have performed three sets of simulations at $w = 0.375$,
0.505838, 0.775, using walks with $100 \le N \le 8000$. 
Results are reported in Tables~\ref{tableratios}, \ref{tableRg},
\ref{tableRe}, and \ref{tableRH}. 
The numerical data that concern $R_g^2$ and $R_e^2$ have been analyzed as 
discussed in Refs.~\CITE{PH-05,CMP-viriale}. Therefore,
if $Q = A_{ge}$, $S_{1,g}$, $S_{2,g}$, $S_{1,e}$, and $S_{2,e}$,
we fit the data to 
\begin{equation}
Q(N,w) \approx Q^* + {a_Q(w)\over N^\Delta} + 
                    {c_Q(w)\over N^{\Delta_{2,\rm eff}}}\; .
\end{equation}
For $\Delta$ we use the best available estimate: 
$\Delta = 0.515\pm 0.007^{+0.010}_{-0.000}$ (Ref.~\CITE{BN-97}).
The term $1/N^{\Delta_{2,\rm eff}}$ should take into account analytic
corrections behaving as $1/N$, and nonanalytic ones of the form
$N^{-2\Delta}$, $N^{-\Delta_2}$ ($\Delta_2$ is the next-to-leading
correction-to-scaling exponent). As discussed in Ref.~\CITE{PH-05},
one can lump all these terms into a single one with exponent 
$\Delta_{2,\rm eff} = 1.0\pm 0.1$. 
The results are reported in Table~\ref{tableris-1}. The reported error 
is the sum of the statistical error and of the systematic one due to 
the uncertainty on $\Delta$ and $\Delta_{2,\rm eff}$. In order to detect
additional corrections to scaling we have repeated the fit several times,
including each time only the data with $N\ge N_{\rm min}$. More precise
estimates can be obtained by using the universality of the ratios of the 
correction-to-scaling amplitudes. As we did in Ref.~\CITE{CMP-viriale},
we can simulteneously analyze two quantities $Q_1$ and $Q_2$, using the fact 
that $a_{Q_1}(w) = r_{12} a_{Q_2}(w)$, with $r_{12}$ independent of $w$.
Thus, one can fit simultaneously all data for $Q_1$ and $Q_2$, 
taking as free parameters $Q_1^*$, $Q_2^*$, 
$a_{Q_1}(w_i)$, $c_{Q_1}(w_i)$, $c_{Q_2}(w_i)$, and $r_{12}$. We have reanalyzed
the data for $S_{1,g}$, $S_{2,g}$, $S_{1,e}$, and $S_{2,e}$ using this method,
taking always as second variable $\hat{R}_g^2/\hat{R}_e^2$. The results are reported
in Table~\ref{tableris-2}. The improvement is quite significant, the errors
decreasing by a factor of 2-3.
We do not observe significant residual scaling corrections and take the result
with $N_{\rm min} = 500$ as our final estimate. Therefore
\begin{eqnarray}
A_{ge}^* &=& 0.15988 \pm 0.00004, \\
S_{1,g}^* &=& -0.3152 \pm 0.0007, \\
S_{2,g}^* &=& -0.087 \pm 0.005,  \\
S_{1,e}^* &=& -0.3853 \pm 0.0007, \\
S_{2,e}^* &=& -0.027 \pm 0.006.
\end{eqnarray}
Note that all density corrections are negative---this is not surprising 
since, by increasing the density, the polymer size decreases. Moreover,
both for $R_{g}^2$ and $R_{e}^2$, the second density correction is 
significantly smaller than the first one. There are in the literature 
several estimates of $A_{ge}^*$ (see Ref.~\CITE{PV-review} for an extensive
list of references). The most precise ones are: 
 $A_{ge}^* = 0.1599 \pm 0.0002$ (Ref.~\CITE{LMS-95}) and 
 $A_{ge}^* = 0.15995 \pm 0.00010$ (Ref.~\CITE{GSS-97}).
They are both in good agreement with our result. We can also 
compare the estimates of $S_{1,g}^*$ and $S_{2,g}^*$ with
those obtained by using one-loop perturbation theory:\cite{Schaefer-99,footnote-Sg}
$S_{1,g}^* \approx - 1.19$ and $S_{2,g}^* \approx 5.74$. 
They both significantly differ from our estimates, indicating that 
the perturbative renormalization group is not accurate (at one-loop order)
for these quantities.

In the analysis of 
quantities involving the hydrodynamic radius we should take into account the
corrections proportional to $N^{\nu - 1} = N^{-0.412}$.
These new corrections are particularly strong
and depend only slightly on $w$, as it can be seen from the results
in Tables~\ref{tableratios}, \ref{tableRH}: 
for all values of $w$, $\hat{R}_g/\hat{R}_H$ and $S_{n,H}$
show the same monotonic behavior, indicating that corrections proportional to
$N^{-\Delta}$ are much smaller than
those proportional to $N^{\nu-1}$ 
(rememeber that the $N^{-\Delta}$ corrections approximately
vanish for $w = 0.505838$ and have opposite sign in the other two cases). 
In order to determine the scaling-limit value of $A_{gH}$, $S_{1,H}$, and 
$S_{2,H}$ we fit the data to 
\begin{equation}
Q(N,w) = Q^* + {a(w)\over N^{1-\nu}} + {c(w)\over N^{\Delta}} \; .
\label{fitA}
\end{equation}
Results are reported in Table \ref{tableris-1} 
for several values of $N_{\rm min}$. 
As before, the error takes into account the uncertainty on $\nu$ and 
$\Delta$. In all cases, the results show small trends with $N_{\rm min}$ 
indicating 
that there are some small additional scaling corrections that are 
not taken into account by the fit Ansatz (\ref{fitA}). We quote as
our final results: 
\begin{eqnarray}
  A_{gH}^* &=& 1.5810 \pm 0.0010 \\
  S_{1,H}^* &=& 0.082 \pm 0.004 \\
  S_{2,H}^* &=& 0.050 \pm 0.020 .
\end{eqnarray}
The error should be large enough to take into account all residual 
systematic corrections.  Note that also in this case density corrections are  
small. 
The result for $A_{gH}^*$ is in good agreement with that obtained 
in Ref.~\CITE{DRSK-02}: $A_{gH}^* = 1.591 \pm 0.007$. 

Finally, we compute the universal scaling-correction coefficients: 
our data are precise enough to provide estimates of $s_{1,g}$, 
$s_{1,e}$, $r_{ge}$, $r_e$, and $r_g$. The first three quantities
can be determined as discussed in Ref.~\CITE{PH-05}. If $Q(N,w)$ 
is a renormalization-group invariant quantity we consider
\begin{equation}
T_Q(N) \equiv {\Psi^*\over Q^*} \left[
  {Q(N,w_1) - Q(N,w_2)\over \Psi(N,w_1) - \Psi(N,w_2)}\right] ,
\label{RQdef}
\end{equation}
which should converge asymptotically to the correction-to-scaling
amplitude with corrections of order $N^{-\Delta_{\rm eff}}$, 
with $\Delta_{\rm eff} = 0.5\pm 0.1$. 
In practice we use the data with $w_1 = 0.375$
and $w_2 = 0.775$ and fit $T_Q(N)$ to 
\begin{equation}
T_Q(N) = a + b N^{-\Delta_{\rm eff}}.
\end{equation}
We find that the correction-to-scaling coefficients are quite small:
\begin{eqnarray}
    r_{ge} &=& -0.0040 \pm 0.0004, \\
    s_{1,g} &=& -0.050 \pm 0.010, \\
    s_{1,e} &=& -0.050 \pm 0.020. 
\end{eqnarray}
As for $s_{2,g}$ and $s_{2,e}$ we only obtain upper bounds: 
$s_{2,g} = -0.1 \pm 0.2$, $|s_{2,e}| \lesssim 0.2$.
In order to determine $r_e$ and 
$r_g$ we first define $Q_R(N) \equiv  \hat{R}^2(2N)/\hat{R}^2(N)$ and then
\begin{equation}
T_{R}(N) \equiv  {2^{-2\nu} \Psi^* \over 1 - 2^\Delta}  \left[
  {Q_R(N,w_1) - Q_R(N,w_2)\over \Psi(2 N,w_1) - \Psi(2 N,w_2)}\right] 
\end{equation}
that converges to $r_e$ or $r_g$, depending on the radius one is considering, 
with corrections of order $N^{-\Delta_{\rm eff}}$. 
We obtain: 
\begin{eqnarray}
   r_g &=&  - 0.36 \pm 0.07, \\
   r_e &=&  - 0.28 \pm 0.08. 
\end{eqnarray}
Finally, if we analyze as before the quantity
\begin{equation}
T_{R_g,R_e}(N) \equiv  {Q_{R_g}(N,w_1) - Q_{R_g}(N,w_2)\over 
            Q_{R_e}(N,w_1) - Q_{R_e}(N,w_2) },
\end{equation}
we obtain a direct estimate of $r_g/r_e$:
\begin{equation}
    r_g/r_e = 1.2 \pm 0.3.
\end{equation}
This is consistent with the estimate 
\begin{equation}
   {r_g\over r_e} = 1 + {r_{ge}\over r_e} = 1.014 \pm 0.004,
\end{equation}
obtained by using $r_{ge}$ and $r_e$. 
These results can be compared with those appearing in the literature. 
Ref.~\CITE{Nickel-91} gives $-0.5\lesssim r_g \lesssim -0.4$ and 
$r_g/r_e = 1.25 \pm 0.04$, while Ref.~\CITE{BN-97} gives $r_e = -0.41\pm 0.01$.
The estimates of $r_g$ and $r_e$ are consistent with ours, while 
that for $r_g/r_e$ differs significantly, given the small error bars.

\section{Conclusions} \label{sec5}

The results of the previous Section allow us to determine the radii in
the dilute regime. We find
\begin{eqnarray}
{R^2_g\over \hat{R}^2_g} &=& 1 - 0.0752 \Phi_p - 0.0050 \Phi^2_p + 
   \cdots + k_\Phi (0.003 \Phi_p + b_g \Phi_p^2 \cdots ),  
\label{expR2g} \\
{R^2_e\over \hat{R}^2_e} &=& 1 - 0.0920 \Phi_p - 0.0015 \Phi^2_p + 
   \cdots + k_\Phi (0.004 \Phi_p + b_e \Phi_p^2 \cdots ), 
\label{expR2e} \\
{\hat{R}_H\over {R}_H} &=& 1 + 0.0195 \Phi_p + 0.003 \Phi^2_p + \cdots ,
\label{expRH} 
\end{eqnarray}
where we have introduced the polymer packing fraction
\begin{equation}
\Phi_p \equiv {4 \pi \hat{R}^3_g\over 3} c = {4 \pi \hat{R}^3_g\over 3} {N_A\over M} \rho,
\end{equation}
$N_A$ is the Avogadro number, $M$ the molar mass of the polymer,
$c$ and $\rho$ the number density and the ponderal concentration respectively.
The constants $b_g$ and $b_e$ depend on $s_{2,g}$ and $s_{2,e}$ and thus 
we only have an upper bound: $|b_g|\lesssim 1\cdot 10^{-3}$, 
$|b_e|\lesssim 2\cdot 10^{-4}$.
The constant $k_\Phi$ takes into account scaling corrections. It
scales as $N^{-\Delta}$, where $N$ is the degree of polymerization, and
therefore vanishes in the scaling limit. It is defined by the small-density
behavior of the osmotic pressure:\cite{CMP-viriale}
\begin{equation}
Z = {\beta \Pi\over c} \approx
     1 + 1.313 \Phi_p + 0.559 \Phi_p^2  - 0.122 \Phi_p^3 + \cdots 
     + k_\Phi (\Phi_p + 1.13 \Phi_p^2 + \cdots). 
\label{Zris}
\end{equation}
Expansions (\ref{expR2g}), (\ref{expR2e}), (\ref{expRH}) are supposed to 
be valid up to $\Phi_p\approx 1$. In this range, the linear term in the 
density dominates, the next-to-leading term being quite small,
at variance with the behavior of the osmotic pressure. Note also
that the density dependence is very small in all cases.
The nonuniversal parameter $k_\Phi$
allows us to take into account the deviations from the scaling limit
due to the finite degree of polymerization (as long as scaling corrections
are small). It can be obtained 
from measurements of the osmotic pressure by using Eq.~(\ref{Zris}). 
Once $k_\Phi$ is known, everything is fixed and the previous expressions give us 
the radii in the whole dilute regime $\Phi_p\lesssim 1$.
Note that corrections to scaling in the radii are much smaller than in
the osmotic pressure. 

\begin{figure}
\centerline{\epsfig{file=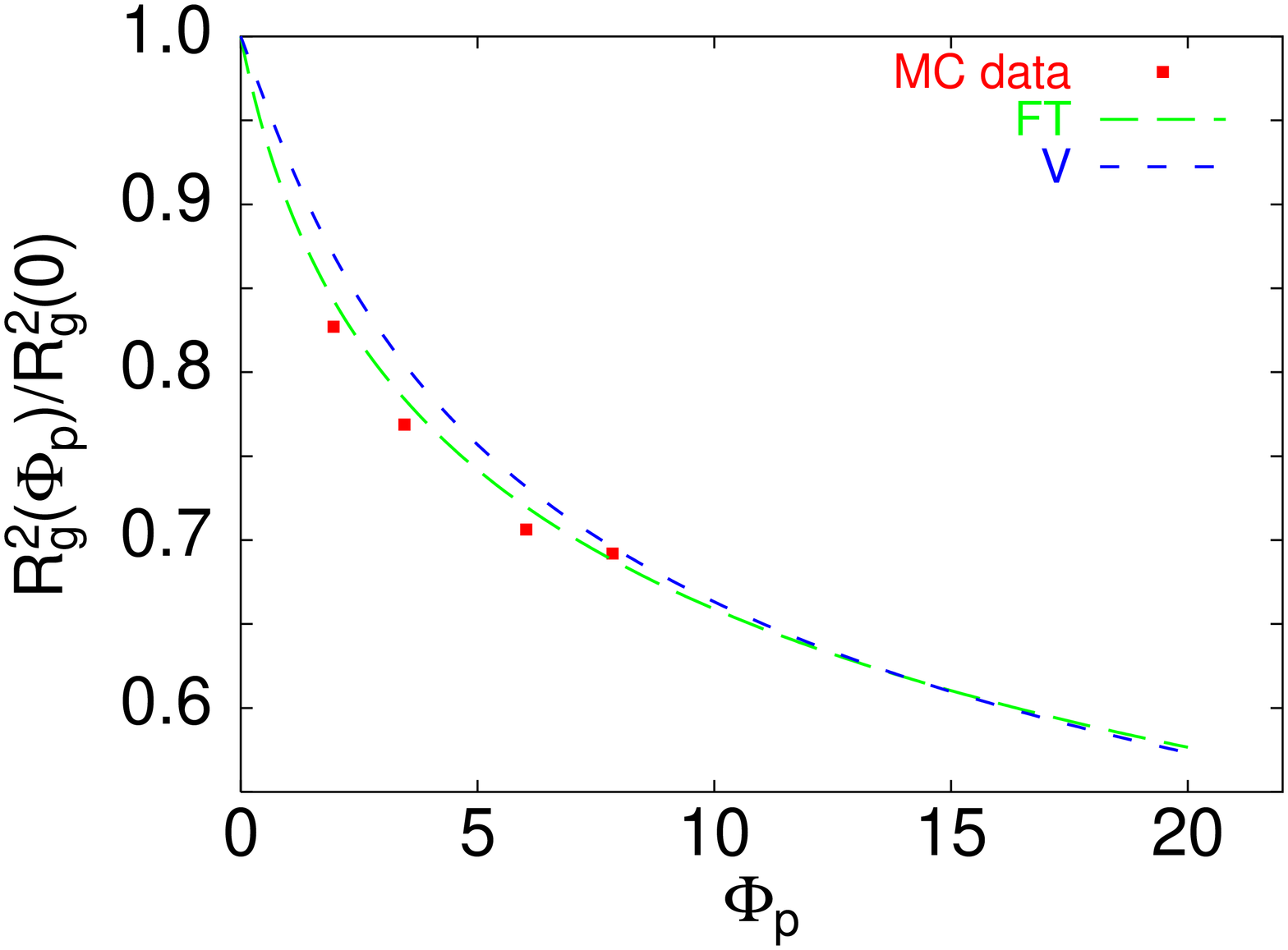,angle=-90,width=16truecm}}
\vspace{1cm}
\caption{Plot of $R^2_g/\hat{R}^2_g$ versus $\Phi_p$.
We report expression (\protect\ref{extra}), ``V", the field-theoretical prediction of 
Ref.~\protect\CITE{ON-83}, ``FT", and the Monte Carlo data of Ref.~\protect\CITE{MBS-00}. 
}
\label{comparison}
\end{figure}

It is wortwhile to try to extrapolate the above-reported expansions outside 
the dilute regime. As in Ref.~\CITE{CMP-viriale} we will use a 
simple interpolation formula that takes into account the 
large-$\Phi_p$ behavior of the radii. For $\Phi_p\to\infty$ we 
have\cite{Schaefer-99}
\begin{equation}
   {R^2\over \hat{R}^2} \sim \Phi_p^{-(2 \nu - 1)/(3 \nu - 1)} 
    \sim \Phi_p^{-0.230},
\label{largePhi}
\end{equation}
which is a consequence of the fact that, at fixed nonvanishing monomer 
concentration, polymers behave as Gaussian coils as $N\to\infty$. 
It is easy to write down interpolation formulas that take into account
Eq.~(\ref{largePhi}) and reproduce the previous expansions for 
$\Phi_p\to \infty$:
\begin{eqnarray}
{R^2_g\over \hat{R}^2_g} &=& 
   (1 + 0.655 \Phi_p + 0.28 \Phi_p^2 )^{-0.115}, \nonumber \\
{R^2_e\over \hat{R}^2_e} &=&
   (1 + 0.801 \Phi_p + 0.37 \Phi_p^2 )^{-0.115}, \nonumber \\
{\hat{R}_H\over {R}_H} &=& 
   (1 + 0.34 \Phi_p + 0.10 \Phi_p^2 )^{0.0574}.
\label{extra}
\end{eqnarray}
We do not attempt to include also scaling corrections here, since 
we have only estimated one single coefficient. 
Of course, these expressions are very precise in the dilute limit.
In order to assess their precision in the semidilute regime we can compare 
with field-theoretical estimates\cite{ON-83,Freed-83,Schaefer-84,Schaefer-99}
and with Monte Carlo results.\cite{MBS-00} In Fig.~\ref{comparison}
we compare expression (\ref{extra}) for the radius of gyration with 
the expression reported in Ref.~\CITE{ON-83}. The two curves are quite close. 
They differ at most by 3.2\% (for $\Phi_p\approx 1.5$) and, as $\Phi_p\to \infty$,
they  give
${R^2_g/ \hat{R}^2_g} \approx 1.18 \Phi_p^{-0.230}$ (Ref.~\CITE{ON-83}) and 
${R^2_g/ \hat{R}^2_g} \approx 1.16 \Phi_p^{-0.230}$ [Eq.~(\ref{extra})].
Note that the largest differences are observed for small values of 
$\Phi_p$ (for $\Phi_p= 0.4$ the curves differ by 2\%) where our expression is 
by construction very accurate. This is related to the fact that,
as observed in Sec.~\ref{sec4}, the low-density coefficients for the radii
are poorly determined by one-loop field-theoretical calculations.
Similar results are obtained by using the expressions reported in 
Ref.~\CITE{Schaefer-99}. For instance, they predict 
${R^2_g/ \hat{R}^2_g} \approx 1.14 \Phi_p^{-0.230}$ [see Eq.~(18.12) in 
Ref.~\CITE{Schaefer-99}]. 
In Fig.~\ref{comparison} we also report some Monte Carlo data 
taken from Ref.~\CITE{MBS-00}. They are somewhat lower 
(differences are less than 5\%) than 
prediction (\ref{extra}) and in better agreeement with the field-theoretical
curve, but they could be affected by finite-size and scaling corrections. 
This comparison shows that 
our expression for the density dependence of $R_g$ is reasonably accurate, 
the error being apparently less than 5\% in the whole semidilute region.
We are not aware of similar field-theoretical results for the other radii,
so that we are unable to estimate the precision of the extrapolation for 
$R_e$ and $R_H$.

\vskip 1truecm

The authors thank Tom Kennedy for providing his efficient simulation code 
for lattice self-avoiding walks.

\begin{table}
\caption{Estimates of the ratios $A_{ge} = \hat{R}^2_g/\hat{R}^2_e$ and 
$A_{gH} = \hat{R}_g/\hat{R}_H$. }
\label{tableratios}
\begin{tabular}{cccc}
$N$ & $w=0.375$ & $w=0.505838$ &  $w=0.775$ \\
\hline
\multicolumn{4}{c}{$A_{ge}$} \\
\hline
100 &0.161771(15) & 0.1609797(95) & 0.159965(14)\\
250 &0.160842(15) & 0.160283(11) & 0.159637(10)\\
500 &0.160487(15) & 0.160079(11) & 0.159612(14)\\
1000 &0.160293(21) & 0.160017(16) & 0.159644(24)\\
2000 &0.160162(22) & 0.159910(15) & 0.159691(27)\\
4000 &0.160067(22) & 0.159917(17) & 0.159753(23)\\
8000 &0.160060(27) & 0.159888(19) & 0.159759(26)\\
\hline
\hline
\multicolumn{4}{c}{$A_{gH}$} \\
\hline
100 & 1.246920(42)  & 1.253683(42)  &   1.263604(39) \\
250 &1.348559(45)  &  1.354758(52)   &  1.363108(43)  \\
500 &1.404691(68)  &  1.410316(71)  & 1.417575(62)  \\
1000 &1.44766(10)  & 1.452355(91)  &  1.458507(99) \\
2000 &1.48008(15)  & 1.48402(15)  &  1.48946(16) \\
4000 &1.50436(23)  & 1.50827(22)  &  1.51251(20) \\
8000 &1.52336(30)  & 1.52607(32)  &  1.52933(32) \
\end{tabular}
\end{table}

\begin{table}
\caption{Estimates of the coefficients $S_{1,g}$ and 
$S_{2,g}$ for the radius of gyration.}
\label{tableRg}
\begin{tabular}{cccc}
$N$ & $w=0.375$ & $w=0.505838$ &  $w=0.775$ \\
\hline
\multicolumn{4}{c}{$S_{1,g}$} \\
\hline
100 &$-$0.28885(58) &$-$0.30737(43)  & $-$0.32464(55)  \\
250 &$-$0.30149(57) &$-$0.31331(46)  & $-$0.32331(45)  \\
500 &$-$0.30663(71) &$-$0.31391(45)  & $-$0.32202(65)  \\
1000 &$-$0.3093(10) &$-$0.31533(72)  & $-$0.3184(11) \\
2000 &$-$0.3121(11) &$-$0.31562(74)  & $-$0.3183(10) \\
4000 &$-$0.3105(11) &$-$0.31412(77)  & $-$0.3161(13) \\
8000 &$-$0.3144(13) &$-$0.31629(87)  & $-$0.3166(14) \\
\hline\hline
\multicolumn{4}{c}{$S_{2,g}$} \\
\hline
100 &$-$0.0451(49) &$-$0.0783(45)  & $-$0.1324(55)  \\
250 &$-$0.0646(59) &$-$0.0905(48)  & $-$0.1234(47)  \\
500 &$-$0.0752(65) &$-$0.0787(51)  &$-$0.1190(72)   \\
1000 &$-$0.071(10) &$-$0.0950(73)  & $-$0.082(11) \\
2000 &$-$0.059(11) &$-$0.0920(78)  & $-$0.105(10) \\
4000 &$-$0.079(11) &$-$0.0883(81)  & $-$0.080(13) \\
8000 &$-$0.099(11) &$-$0.0896(89)  & $-$0.080(12) \\
\end{tabular}
\end{table}

\begin{table}
\caption{Estimates of the coefficients $S_{1,e}$ and 
$S_{2,e}$ for the end-to-end distance.}
\label{tableRe}
\begin{tabular}{cccc}
$N$ & $w=0.375$ & $w=0.505838$ &  $w=0.775$ \\
\hline
\multicolumn{4}{c}{$S_{1,e}$} \\
\hline
100 & $-$0.35414(79) &$-$0.37943(55)  & $-$0.40439(68) \\
250 & $-$0.36735(78) & $-$0.38421(58)  & $-$0.39891(56)  \\
500 & $-$0.37364(85) & $-$0.38442(56) &  $-$0.39570(91) \\
1000 &$-$0.3780(13)  & $-$0.38556(85) & $-$0.3911(11)   \\
2000 &$-$0.3814(13)  &$-$0.38608(91)  & $-$0.3891(13) \\
4000 & $-$0.3801(13) &$-$0.38397(92)  &$-$0.3866(14)  \\
8000 &$-$0.3848(16)  &$-$0.38609(97)  & $-$0.3878(16) \\
\hline\hline
\multicolumn{4}{c}{$S_{2,e}$} \\
\hline
100  & 0.0050(67)  &$-$0.0271(62)  & $-$0.0862(86) \\
250  & $-$0.0058(86) &$-$0.0364(59)  &$-$0.0694(65)  \\
500  & $-$0.0139(82) &$-$0.0225(69)  & $-$0.062(10) \\
1000 & $-$0.009(13) &$-$0.040(10) & $-$0.025(13) \\
2000 & $-$0.006(15) & $-$0.042(10) & $-$0.038(14) \\
4000 & $-$0.040(14) &$-$0.018(10)& $-$0.008(14) \\
8000 &$-$0.039(16)  &$-$0.035(11) & $-$0.025(14) \\
\end{tabular}
\end{table}

\begin{table}
\caption{Estimates of the coefficients $S_{1,H}$ and $S_{2,H}$ 
for the hydrodynamic radius.}
\begin{tabular}{cccc}
$N$ & $w=0.375$ & $w=0.505838$ &  $w=0.775$ \\
\hline
\multicolumn{4}{c}{$S_{1,H}$} \\
\hline
100  &0.09295(19)  &0.09747(18)  &  0.10014(19)  \\
250  &0.09239(20)  &0.09491(21)  &  0.09578(20)  \\
500  &0.09121(25)  &0.09221(28)  &  0.09313(20) \\
1000 &0.08991(39)  &0.09031(40)  &  0.09097(34)  \\
2000 &0.08739(48)  &0.08874(53)  &  0.08768(50)  \\
4000 &0.08708(68)  &0.08822(53)  &  0.08747(59)  \\
8000 &0.08631(84)  &0.08612(73)  &  0.08564(69)  \\
\hline\hline
\multicolumn{4}{c}{$S_{2,H}$} \\
\hline
100 &0.0668(19)   & 0.0851(21)  & 0.0999(24)   \\
250  &0.0731(23)   & 0.0826(22) & 0.0967(23)  \\
500  &0.0782(30)   &0.0822(27)  & 0.0902(29)   \\
1000 &0.0811(42)  & 0.0792(40) & 0.0929(41) \\
2000 &0.0730(50)   & 0.0911(62) &0.0747(62)  \\
4000 & 0.0842(78)  & 0.0819(72) & 0.0748(81)  \\
8000 &0.0773(97)   &0.0738(94)   &0.058(10)  \\
\end{tabular}
\label{tableRH}
\end{table}

\begin{table}
\caption{Fit results for several values of $N_{\rm min}$.}
\label{tableris-1}
\begin{tabular}{ccccc}
& $N_{\rm min} = 100$  & $N_{\rm min} = 250$ & $N_{\rm min} = 500$ &
  $N_{\rm min} = 1000$  \\
\hline
$A_{ge}^*$ & 
 0.15991(4)  &   0.15989(4)&    0.15988(4) &   0.15995(7) \\
$A_{gH}^*$ & 
 1.5831(4) & 1.5819(4) & 1.5811(7) & 1.5806(13) \\
$S_{1,g}^*$ &
  $-$0.3144(8)   &$-$0.3150(10)& $-$0.3152(15)& $-$0.3165(30) \\
$S_{2,g}^*$ &
   $-$0.082(6)    &$-$0.090(9)  & $-$0.093(15) & $-$0.095(27)  \\
$S_{1,e}^*$ &
  $-$0.3846(9) & $-$0.3851(11) & $-$0.3851(18) & $-$0.3873(36) \\
$S_{2,e}^*$ &
  $-$0.023(8) & $-$0.031(12) & $-$0.032(19) & $-$0.036(36) \\
$S_{1,H}^*$ &
   0.0809(7) & 0.0815(10) & 0.0821(18) & 0.0866(36) \\
$S_{2,H}^*$ & 
   0.068(7) & 0.059(12) & 0.044(22) & 0.050(44) \\
\end{tabular}
\end{table}

\begin{table}
\caption{Estimates of the density corrections to 
$R_g$ and $R_e$. Combined fits with data for $A_{ge}$.}
\label{tableris-2}
\begin{tabular}{ccccc}
& $N_{\rm min} = 100$  & $N_{\rm min} = 250$ & $N_{\rm min} = 500$ &
  $N_{\rm min} = 1000$  \\
\hline
$S_{1,g}^*$ &
  $-$0.3149(9) & $-$0.3151(6) & $-$0.3152(7) & $-$0.3143(13) \\
$S_{2,g}^*$ &
  $-$0.087(5) & $-$0.086(4) & $-$0.087(5) & $-$0.087(9) \\
$S_{1,e}^*$ &
  $-$0.3846(13) & $-$0.3852(8) & $-$0.3853(7) & $-$0.3843(14) \\
$S_{2,e}^*$ &
  $-$0.028(8) & $-$0.027(5) & $-$0.027(6) & $-$0.034(15) \\
\end{tabular}
\end{table}

\begin{thebibliography}{199}
\bibitem{footPsi}
The interpenetration radius $\Psi$ 
is determined from the low-density behavior of
the osmotic pressure $\Pi$. If $\Pi/(R T\rho) = 1/M + B_2 \rho + O(\rho^2)$ 
($M$ is the molar mass of the polymer, $\rho$ the ponderal concentration, and 
$T$ the absolute temperature), then we define
$\Psi \equiv  2 (4\pi)^{-3/2} M^2 B_2 R^{-3}_g/N_A$, where $N_A$ is the 
Avogradro number.
\end{thebibliography}
\end{document}